\documentclass[twocolumn,aps,amsmath,amssymb]{revtex4}
\usepackage[top=2.5cm]{}
\usepackage{graphicx}
\usepackage{mathrsfs,amsmath}
\usepackage{bm}

\newcommand{\ti}{t_{0}}

\begin{document}

\title{Eddy diffusivities  of inertial particles in random Gaussian flows}
\author{S. Boi$^{1,2}$, A. Mazzino$^{1,2,3}$ and  P. Muratore-Ginanneschi$^4$}
\affiliation{$^1$DICCA, University of Genova, Via Montallegro 1, 16145 Genova, Italy\\
$^2$ INFN,   Genova Section, Via Dodecaneso 33, 16146 Genova, Italy\\
$^3$ CINFAI Consortium, Genova Section, Via Montallegro 1, 16146 Genova, Italy\\
$^4$Department of Mathematics and Statistics, University of Helsinki, Gustaf Haellstroemin katu 2b, Helsinki
}
\date{\today}

\begin{abstract} 
We investigate the large-scale transport of inertial  particles. We 
derive explicit analytic expressions for the eddy diffusivities 
for generic Stokes times.  These latter expressions are exact for any 
shear  flow while they  correspond to  the  leading contribution either
in the deviation from the shear flow geometry or in the P\'eclet  number of general
random Gaussian velocity fields.   
Our explicit expressions allow us to investigate the role  
of inertia for such  a class  of flows and to  make exact links with  
the analogous transport problem for tracer particles.
\end{abstract}


\maketitle


Understanding the role  of particle inertia on the
late-time dispersion process is a problem of paramount importance in a
variety of  situations, mainly  related to geophysics  and atmospheric
sciences. Airborne particulate  matter in the atmosphere  has indeed a
well-recognized role  for the  Earth's climate  system because  of its
effect  on  global  radiative   budget  by  scattering  and  absorbing
long-wave  and  short-wave radiation  \cite{IPCC}.   For  the sake  of
example, one of  the most intriguing issue in this  context is related
to the evidence of anomalous large fluctuations in the residence times
of mineral dust  observed in different experiments carried  out in the
atmosphere \cite{Denjean2016}.

Those  observations  naturally lead  to  the  idea that  settling  and
dispersion of  inertial particles, both contributing  to the residence
time of particles in the  atmosphere, crucially depend on the peculiar
properties of the carrier flow encountered in the specific experiment.
For  the  gravitational  settling,  this  question  was  addressed  in
Ref.~\cite{martins2008}.  It turned  out that the value  of the Stokes
number  alone, $St$,  directly related  to the  particle size,  is not
sufficient  to argue  if the  sedimentation is  faster or  slower with
respect to what  happens in still fluid. With minor  variations of the
carrier  flow, for  a given  $St$, it  has been  shown that  either an
increase  or a  reduction of  the falling  velocity are  possible thus
affecting in a different way the particle residence time in the fluid.

Our  aim here  is to  shed some  light on  how dispersion  of inertial
particles does depend on relevant  properties of the turbulent carrier
flow. Our  focus will be  on the  late-time evolution of  the particle
dynamics,  a regime  fully  described in  terms of  eddy-diffusivities
\cite{frischeddy,F95,mamamu2012}.   Our  main  question  can  be  thus
rephrased in terms of the behavior  of the eddy diffusivity by varying
some  relevant features  of the  carrier flow  (e.g. the  form of  its
auto-correlation  function), for  a given  inertia of  the particle.\\
This  analysis for  generic  carrier  flows is  a  task of  formidable
difficulty  and  forces to  the  exploitation  of numerical  approaches
which,  however, make  it difficult  to isolate  simple mechanisms  on
large-scale transport induced by inertia.  To overcome the problem, we
decided  to focus  on  simple  flow field  where  the  problem can  be
entirely grasped via analytic (or perturbative) techniques. As we will
see,  shear flows  are natural  candidates to  allow one  the
analytic treatment of large-scale transport.

Let us considered the well-known model \cite{MR83,G83} 
for  transport of heavy particles in $d$-spatial dimensions 
by an incompressible carrier flow $\boldsymbol{u}(\bm{\xi}(t),t)$:
\begin{eqnarray}
\label{sf:sdepos}
\begin{array}{l}
\mathrm{d}\boldsymbol{\xi}(t)=\,\boldsymbol{v}(t)\,\mathrm{d}t 
\\[0.1cm]
\mathrm{d}\boldsymbol{v}(t)=-\,\left(\dfrac{\boldsymbol{v}(t)
-\boldsymbol{u}(\boldsymbol{\xi}(t),t)}{\tau}\right)\,\mathrm{d}t
+\dfrac{\sqrt{2\,D_0}}{\tau}
\,\mathrm{d}\boldsymbol{\omega}(t)
\end{array}
\end{eqnarray}  
Here $\boldsymbol{v}$  denotes the  particle velocity,
$\bm{\xi}$   its   trajectory,  $\tau$   is   the   Stokes  time. Finally,
$\boldsymbol{\omega}$  denotes   a  standard   $d$-dimensional  Wiener
process  \cite{Jacobs}.    Increments  $\mathrm{d}\boldsymbol{\omega}$ 
coupled to (\ref{sf:sdepos}) by a constant molecular diffusivity $D_0$
model, as customary, fast scale  chaotic forces acting on the inertial
particle acceleration \cite{R88}. \\  
To start with,  we assume that  the carrier flow is a shear   
\begin{eqnarray}
\boldsymbol{u}(\boldsymbol{x},t)=u(x_{2},\dots,x_{d},t)\,\boldsymbol{e}_{1}
\nonumber
\end{eqnarray}
where $\boldsymbol{e}_{1}=(1,0,\dots,0)$ is the constant unit vector pointing along the
first axis. This simple geometry readily enforces the incompressibility
condition. We  also assume  that $u$ is  a stationary
and  homogeneous  Gaussian  random  field  with  mean  and  covariance
specified by
\begin{eqnarray}
\label{cs}
\begin{array}{l}
\langle u(x_2,\dots,x_d,t) \rangle =0
\\[0.1cm]
\langle u(x_2,\dots,x_d,t)\, u(0,\dots,0,0) \rangle= B(x_2,\dots,x_d,|t|)
\end{array}
\end{eqnarray}
It is worth stressing that we assume that the Eulerian statistics of the 
carrier flow is independent from the Wiener process driving (\ref{sf:sdepos}).
For a shear flow, (\ref{sf:sdepos}) is integrable
by elementary techniques. We find
\begin{subequations}
\label{sol:n}
\begin{eqnarray}
\label{ppp}
v_{n}(t)=
e^{-\frac{t-t_{o}}{\tau}}\,{v}_{n}(\ti)+\frac{\sqrt{2\,D_0}}{\tau}\int\limits_{t_{o}}^{t}\mathrm{d}{\omega}_{n}(s)\,e^{-\frac{t-s}{\tau}}
\end{eqnarray}
\begin{eqnarray}
\label{POSITIONTRAN}
\lefteqn{\hspace{-1.25cm}
\xi_{n}(t)=\xi_{n}(\ti)+\tau  (1-e^{-\frac{t-t_{o}}{\tau}}){v}_{n}(\ti)
}
\nonumber\\&&
+\sqrt{2\,D_0}\int\limits_{t_{o}}^{t}\mathrm{d}{\omega}_{n}(s)\,(1-e^{-\frac{t-s}{\tau}})
\end{eqnarray}
\end{subequations}
for $n\neq1$, and
\begin{subequations}
\label{sol:1}
\begin{eqnarray}
\label{vel1}
\lefteqn{\hspace{-0.25cm}
v_{1}(t)
=e^{-\frac{t-t_{o}}{\tau}}\,{v}_{1}(\ti)+\frac{\sqrt{2\,D_0}}{\tau}\int\limits_{t_{o}}^{t}\mathrm{d}{\omega}_{1}(s)\,e^{-\frac{t-s}{\tau}}
}
\nonumber\\&&
+\frac{1}{\tau}\int\limits_{t_{o}}^{t}\mathrm{d} s\, u(\xi_{2}(s),\dots,\xi_{d}(s),s)\,e^{-\frac{t-s}{\tau}}
\end{eqnarray}
\begin{eqnarray}
\label{qqq}
\lefteqn{
\xi_{1}(t)={\xi}_{1}(\ti)+\tau {v}_{1}(\ti)(1-e^{-\frac{t-t_{o}}{\tau}})}
\nonumber\\
&&+\sqrt{2\,D_0}\int\limits_{t_{o}}^{t}\mathrm{d}{\omega}_{1}(s)\,(1-e^{-\frac{t-s}{\tau}})
\nonumber\\
&&+\int\limits_{t_{o}}^{t}\mathrm{d} s\, u(\xi_{2}(s),\dots,\xi_{d}(s),s)\,(1-e^{-\frac{t-s}{\tau}})
\end{eqnarray}
\end{subequations}
for n=1. The stochastic integrals appearing in (\ref{sol:1}), 
(\ref{sol:n}) can be interpreted as the limit of usual Riemann sums owing
to the additive nature of the noise.

A relevant indicator of the  dispersion properties of a single particle 
trajectory is the effective diffusion tensor defined as 
\begin{eqnarray}
\label{DEF}
\mathsf{D}_{l n}^{\mathrm{eff}}=\lim_{t\uparrow \infty}\frac{\langle \xi_{l}(t)\,\xi_{n}(t) \rangle
-\langle \xi_{l}(t)\rangle\,\langle\xi_{n}(t) \rangle}{2\,(t-t_{0})}
\nonumber
\end{eqnarray}
or, equivalently, by a straightforward application of de l'H\^opital rule 
\begin{eqnarray}
\label{Def}
\mathsf{D}_{l n}^{\mathrm{eff}}=\lim_{t\uparrow \infty}\frac{\langle v_{l}(t)\,\xi_{n}(t) \rangle
-\langle v_{l}(t)\rangle\,\langle\xi_{n}(t)\rangle +l \leftrightarrow n}{2}
\end{eqnarray}
Inspection of (\ref{sol:1}), (\ref{sol:n}) readily shows that the 
only non-vanishing elements of the effective diffusion tensor are diagonal
and are specified by the correlations  $\langle \xi_{n}(t)\,\xi_{n}(t) \rangle$
$n=1,\dots,d$ (here and in the following the Einstein convention on repeated indexes
is not adopted).
A straightforward calculation yields the explicit value of the correlations
\begin{eqnarray}
\label{nneq1}
\lefteqn{\mathrm{D}_{n n}^{\mathrm{eff}}=
\lim_{t\uparrow\infty} \langle{v}_{n}(t){\xi}_{n}(t)\rangle}
\nonumber\\
&&=\frac{2\,D_0}{\tau}\int\limits_{0}^{\infty}ds \,(1-e^{-\frac{s}{\tau}})\,e^{-\frac{s}{\tau}}=D_{0}
\end{eqnarray}
 for $n\neq 1$. The carrier flow appears only in the 
correlation function for $n=1$. We find
 \begin{eqnarray}
 \label{n1}
\lefteqn{
\lim_{t\uparrow\infty}\langle \xi_{1}(t) v_{1}(t)\rangle=D_0
+\lim_{t\uparrow\infty}
}
\\&&\hspace{-0.5cm}
\times\int\limits_{(t_{0},t)^{2}} \hspace{-0.2cm}\mathrm{d}s\mathrm{d}s^{\prime}\,
\frac{e^{-\frac{t-s}{\tau}}(1-e^{-\frac{t-s'}{\tau}})
\langle u(\boldsymbol{\eta}(s,t_0),s)\,u(\boldsymbol{\eta}(s^{\prime},t_0),s^{\prime})\rangle}{\tau}\nonumber
\end{eqnarray}
 for $\boldsymbol{\eta}(s,t_0)=(\xi_{2}(s),\dots,\xi_{d}(s))$ and  $ \xi_{i} (t)$ $i=2,\dots,d$ given by Eq.~(\ref{POSITIONTRAN}).  It is worth observing that the explicit
 dependence on $t_0$ in
 (\ref{n1}) actually disappears due to the limit $t\uparrow\infty$.  Without loss of generality we can thus assume  $t_0 = -\infty$ in  (\ref{n1}) in order to obtain simpler expressions.
The integrand in (\ref{n1}) is amenable to a more explicit 
form, if we represent the Eulerian correlation function $B$ 
of the carrier flow, defined in (\ref{cs}), in terms of its 
Fourier representation. In such a case, the average over the
Eulerian statistics of the carrier flow and the Lagrangian
statistics of the first $d-1$ coordinates of the inertial 
particle factor out as
\begin{eqnarray}
\label{correlazione}
\lefteqn{
\langle u(\boldsymbol{\eta}(s,t_0),s)\,u(\boldsymbol{\eta}(s^{\prime},t_0),s^{\prime})\rangle
}\nonumber\\&&
= \int\limits_{\mathbb{R}^{d-1}}\frac{\mathrm{d}^{d-1}\boldsymbol{k}}{(2\,\pi)^{d-1}}\check{\mathsf{B}}(\boldsymbol{k},|s-s'|)
\langle e^{\imath\, \boldsymbol{k}\cdot (\boldsymbol{\eta}(s,t_0)-\boldsymbol{\eta}(s^{\prime},t_0)}\rangle
\end{eqnarray}
After some tedious yet elementary manipulations involving Gaussian 
integration on the Wiener process
and changes of variables in the plane $(s,s')$, 
we obtain
\begin{eqnarray}
\lefteqn{
\mathrm{D}_{11}^{\mathrm{eff}}=D_0+
}
\nonumber\\&&
\hspace{-0.5cm}
\int\limits_{\mathbb{R}^{d-1}}\frac{\mathrm{d}^{d-1}\boldsymbol{k}}{(2\,\pi)^{d-1}}
\int\limits_{0}^{\infty}\mathrm{d}t\,
e^{-D_{0} \|\boldsymbol k\|^2\left[t-\,\tau \, \left( 1-e^{-\frac{t}{\tau }}\right)\right]} 
\check{\mathsf{B}}(\boldsymbol{k},t)
\label{D11}
\end{eqnarray}
We therefore see that all the dynamically non trivial information 
is encoded in the isotropic component of the effective diffusion tensor
\begin{eqnarray}
\label{eddyfinale}
\lefteqn{\hspace{-0.3cm}
D^{\mathrm{eff}}=\frac{1}{d}\sum_{n=1}^{d}\mathrm{D}_{n n}^{\mathrm{eff}}=
D_0+}
\nonumber\\&&
\hspace{-0.3cm}
\int\limits_{\mathbb{R}^{d-1}}\frac{\mathrm{d}^{d-1}\boldsymbol{k}}{(2\,\pi)^{d-1}}
\int\limits_{0}^{\infty}\mathrm{d}t\,
e^{-D_{0} \|\boldsymbol k\|^2\left[t-\,\tau \, \left( 1-e^{-\frac{t}{\tau }}\right)\right]} 
\frac{\check{\mathsf{B}}(\boldsymbol{k},t)}{d}
\end{eqnarray}
We emphasize that (\ref{D11}) and the resulting expression for the 
isotropic component of the effective diffusion tensor are exact results. 
There are several reasons why these simple results are interesting.
To start with we notice that although derived for the highly stylized case 
of shear flow, they continue to hold in suitable asymptotic senses 
for much general classes of carrier flows.
Namely, our final result for the isotropic component $D^{\mathrm{eff}}$
of the effective diffusion tensor coincides, with the one
for tracer particles with colored noise derived in \cite{MaCa99}. 

More generally, $D^{\mathrm{eff}}$ admits the same expression if we compute the eddy diffusivity 
tensor in an infra-red perturbative expansion in the coupling of the carrier flow. 
The logic of the calculation is the same as in \cite{MazzVerg} but applied to inertial 
rather than Lagrangian particles. First, we couch (\ref{sf:sdepos}) into the equivalent 
integral form
\begin{eqnarray}
\label{irsol}
\begin{array}{l}
\boldsymbol{v}(t)=\boldsymbol{v}^{(0)}(t)
+\dfrac{1}{\tau}\int\limits_{t_{o}}^{t}\mathrm{d} s\, \boldsymbol{u}(\boldsymbol{\xi}(s),s)\,e^{-\frac{t-s}{\tau}}
\\[0.3cm]
\boldsymbol{\xi}(t)=\boldsymbol{\xi}^{(0)}(t)
+\int\limits_{t_{o}}^{t}\mathrm{d} s\, \boldsymbol{u}(\boldsymbol{\xi}(s),s)\,(1-e^{-\frac{t-s}{\tau}})
\end{array}
\end{eqnarray}
where now $\boldsymbol{\xi}^{(0)}(t)$, $\boldsymbol{v}^{(0)}(t)$ are Gaussian processes with components 
(\ref{sol:n}) but for $n=1,\dots,d$.  Let us assume the carrier flow to be an 
incompressible Gaussian random field with homogeneous and stationary 
statistics
\begin{eqnarray}
\begin{array}{l}
\langle\boldsymbol{u}(\boldsymbol{x},t)\rangle=0
\\[0.3cm]
\langle\,u_{l}(\boldsymbol{x},t)\,u_{n}(\boldsymbol{0},t)\rangle=\mathsf{B}_{l n}(\boldsymbol{x},|t|)
\end{array}
\nonumber
\end{eqnarray}
Upon inserting (\ref{irsol}) into (\ref{Def}) and retaining
the leading order in $\boldsymbol{u}$ (corresponding either to small $\mathsf{B}$ compared to
$(D_0/L)^2$ -- $L$ being a characteristic length-scale of the flow -- or neglecting
small deviations from the shear-flow geometry), we obtain
\begin{eqnarray}
\lefteqn{
\langle\boldsymbol{v}(t)\cdot\boldsymbol{\xi}(t)\rangle=\langle\boldsymbol{v}^{(0)}(t)\cdot\boldsymbol{\xi}^{(0)}(t)\rangle
}
\nonumber\\&&
\hspace{-0.1cm}
+\tau\int\limits_{t_{0}}^{t}\,\mathrm{d}s_{1}\int\limits_{t_{0}}^{s_{1}}\mathrm{d}s_{2}\,(1-e^{-\frac{t-s_{1}}{\tau}})(1-e^{-\frac{s_{1}-s_{2}}{\tau}})
C_{1}
\nonumber\\&&
+\int\limits_{t_{0}}^{t}\mathrm{d}s_{1}\int\limits_{t_{0}}^{s_{1}}\mathrm{d}s_{2}\,e^{-\frac{t-s_{1}}{\tau}}(1-e^{-\frac{s_{1}-s_{2}}{\tau}})C_{2}
\nonumber\\&&
+\int\limits_{(t_{0},t)^{2}}\mathrm{d}s_{1}\mathrm{d}s_{2}\,(1-e^{-\frac{t-s_{1}}{\tau}})e^{-\frac{t-s_{2}}{\tau}}C_{3}+\dots
\nonumber
\end{eqnarray}
where the $\dots$ symbol stands for higher order terms and
\begin{eqnarray}
\begin{array}{l}
C_{1}=\langle\boldsymbol{v}^{(0)}(t)\cdot(\boldsymbol{u}(\boldsymbol{\xi}^{(0)}(s^{\prime}),s^{\prime})
\cdot\partial_{\boldsymbol{\xi}^{(0)}_{s}})\boldsymbol{u}(\boldsymbol{\xi}^{(0)}(s),s)\rangle
\\[0.2cm]
C_{2}=\langle\boldsymbol{\xi}^{(0)}(t)\cdot(\boldsymbol{u}(\boldsymbol{\xi}^{(0)}(s^{\prime}),s^{\prime})\cdot\partial_{\boldsymbol{\xi}^{(0)}_{s}})
\boldsymbol{u}(\boldsymbol{\xi}^{(0)}(s),s)
\\[0.2cm]
C_{3}=\langle\boldsymbol{u}(\boldsymbol{\xi}^{(0)}_{s^{\prime}},s^{\prime})\cdot\boldsymbol{u}(\boldsymbol{\xi}^{(0)}_{s},s)\rangle
\end{array}
\nonumber
\end{eqnarray} 
If we now invoke the incompressible carrier flow hypothesis we see that $C_{1}$
and $C_{2}$ vanish and that
\begin{eqnarray}
\hspace{-0.1cm}C_{3}\hspace{-0.05cm}=\hspace{-0.1cm}
\int\limits_{\mathbb{R}^{d}}\hspace{-0.05cm}
\frac{\mathrm{d}^{d}\boldsymbol{k}}{(2\,\pi)^{d}} \hspace{-0.05cm}
\sum\limits_{n=1}^{d}\check{\mathsf{B}}_{n n}(\boldsymbol{k},|s-s^{\prime}|) 
\langle\,e^{\imath\boldsymbol{k}\cdot(\boldsymbol{\xi}^{(0)}(s)-\boldsymbol{\xi}^{(0)}(s^{\prime}))}\rangle
\label{gf}
\end{eqnarray}
which coincides with (\ref{correlazione}) in one extra dimension once we identify the trace
of the Fourier transform of the correlation tensor $\mathsf{B}_{l n}$.

After having made the case for the general relevance for the expression of
$D^{\mathrm{eff}}$ we now turn to analyze its behavior as function of the Stokes
number and the characteristic time scale of the carrier flow.

Let us first consider the limit of small $D_0$. 
This would make the resulting integrals easier to manage and to carry out. 
A first order expansion on $D_0$ carried out on  Eq.~(\ref{eddyfinale}) gives:
\begin{eqnarray}
\label{EddyInerAppr}
&\mathrm{D}^{\mathrm{ef}}&=D_0+\frac{1}{d }\int\frac{\mathrm{d}^{d-1}\boldsymbol{k}}{(2\,\pi)^{d-1}}\int_0^\infty\mathrm{d}t\operatorname{tr}
\check{\mathsf{B}}(\boldsymbol{k},t)\nonumber\\
&\times &\,
\,\, \left(1-D_0\|\boldsymbol{k}\|^{2}\left(t-\, \tau \, \left( 1-e^{-\frac{t}{\tau }}\right)\right) \,\right)\,+\dots\nonumber
\end{eqnarray}
or, in physical space:
\begin{eqnarray}
\lefteqn{
D^{\mathrm{eff}}=D_0 +\frac{1}{d }\int_0^\infty \mathrm{d}t 
\langle \boldsymbol u(\boldsymbol x, t) \cdot \boldsymbol u(\boldsymbol x,0)\rangle
}
\nonumber\\&&
\hspace{-0.3cm}
-D_0\sum_{\alpha,\beta=1}^{d} \int\limits_0^\infty \mathrm{d}t\frac{t-\, \tau \, ( 1-e^{-\frac{t}{\tau }})}{d}\,
\langle[\partial_{\alpha} u_\beta(\boldsymbol x, t)]  [\partial_{\alpha} u_{\beta}(\boldsymbol x, 0)]\rangle  
\nonumber\\&&+\dots
\end{eqnarray}
For $\tau\to 0$, the  limit of vanishing inertia easily follows:
\begin{eqnarray}
&&D^{\mathrm{eff}}\longrightarrow_{\tau\to 0}D_0+\frac{1}{d }\int_0^\infty \mathrm{d}t\,
\langle \boldsymbol u(\boldsymbol x, 0) \cdot \boldsymbol u(\boldsymbol x, t)\rangle\nonumber\\
&& -\frac{D_0}{d}\sum_{\alpha,\beta=1}^{d}  \int_0^\infty \mathrm{d}t\,t\, \langle[\partial_{\alpha} u_\beta(\boldsymbol x, 0)]  [\partial_{\alpha} u_{\beta}(\boldsymbol x, t)]\rangle  \nonumber\\
&&+\dots
\end{eqnarray}
which corresponds to the result reported in \cite{MazzVerg}.

Returning to the heavy particle case, in order to further simplify 
the expression for the eddy diffusivity, 
let us focus on a 2D carrier flow with a single wave-number $\boldsymbol k_0$. 
The correlation function we consider is \cite{Antonov}:

\begin{eqnarray}
\label{corr}
&\operatorname{tr}\check{\mathsf{B}}(\boldsymbol{k},|t_{}|)=(2\pi)^{d-1}  E(\boldsymbol k_0) e^{-\frac{|t|}{T_c}} \cos (\Omega t)\nonumber\\
&\times [\delta(\boldsymbol k-\boldsymbol k_0)+\delta(\boldsymbol k+\boldsymbol k_0)]
\end{eqnarray}

$E(\boldsymbol k_0)$ being the turbulent kinetic energy associated to
the wave-number. In principle, the decay time $T_c$ would
depend on $\boldsymbol k$ itself, tipically like $ 1/\|\boldsymbol k\|$ or $ 1/\|\boldsymbol k\|^2$ 
\cite{Kaneda, BoiMazzLac, dissip}. However, since
we are considering a single wave-number flow, we can consider it as a constant.
We can now nondimensionalize our system by setting
$\boldsymbol k_0 =T_c= 1$  and dimensionless, as to have the Stokes
number $St=\tau$. By plugging Eq. (\ref{corr}) into 
Eq. (\ref{EddyInerAppr}), one obtains :
\begin{eqnarray}
\label{eddydiffD0}
\begin{array}{l}
D^{\mathrm{eff}}=D_0+E(\boldsymbol k_0)\left[ \dfrac{1}{d}\dfrac{2}{1+\Omega^2}+\dfrac{D_0}{ d}\,\mathcal K \right]
\\[0.3cm]
\mathcal K=\dfrac{2(1+\text{St})}{1+\Omega^2}-\dfrac{4}{(1+\Omega^2)^2}
+\dfrac{2\text{St}^2(1+\text{St})^2}{(1+\text{St}(2+\text{St}+\text{St}\Omega^2))^2}
\\[0.3cm]
\hspace{0.6cm}
+\dfrac{\text{St}^2(2+\text{St})}{4+\text{St}(4+\text{St}+\text{St}\Omega^2)}
-\dfrac{\text{St}^2(4+3\text{St})}{1+\text{St}(2+\text{St}+\text{St}\Omega^2)}
\end{array} 
\nonumber
\end{eqnarray} 
The above expression is uniform  in St. Indeed, it is a
continuous  function  of St$\in[0,+\infty)$,  and  it  tends to  0  as
St$\to+\infty$ $\forall \Omega$,  then it is limited for  any St. This
means that the  perturbation expansion at first order in  $D_0$ can be
used   for   any   value   of   St.    However,   note   that,   since
$\text{max}|\mathcal K|\leq1$, we have a  constraint on $D_0$ in order
to    have    a    uniform   perturbation    expansion,    which    is
$D_0\ll\frac{2}{(1+\Omega^2)} $.

\begin{figure}
\includegraphics[trim=0 0 0 0,clip,width=6cm,height=6cm]{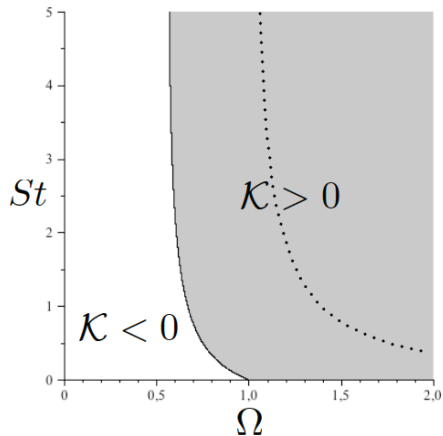}
\caption{The sign of $\mathcal K$ in the $St-\Omega$ plane. Gray corresponds to $\mathcal K >0$;
white to $\mathcal K <0$. The dotted line separates the region on its left, corresponding to transport enhancement due to inertia, from that on its right relative to transport reduction.}
\label{FigNuova}
\end{figure}

  The term $\mathcal K$ can  be either positive or negative, depending
on the  importance of negative  correlated regions in  the correlation
function   (\ref{corr}).     This   fact   can   be    detected   from
Fig.~\ref{FigNuova}  where the  regions inside  which $\mathcal  K$ is
positive (gray  region) and negative  (white region) are shown  in the
plane $St-\Omega$.  It is worth  recalling that, for the  tracer case,
the condition for having $\mathcal K >0$ is simply $\Omega>1$.
The presence of inertia thus causes a change of the sign of
$\mathcal K$ from negative to positive in a subset of the  $St-\Omega$ plane. In this region
inertia thus plays to increase transport with respect to the tracer case.  The region where
transport is enhanced with respect to the tracer case  actually extends up to the dotted line.
To observe a reduction of transport, the Stokes time has thus to be sufficiently large. Larger and larger values are required  for increasing $\Omega$.

\begin{figure}[b]
\includegraphics[trim=0 0 0 0,clip,width=6cm, height=5cm]{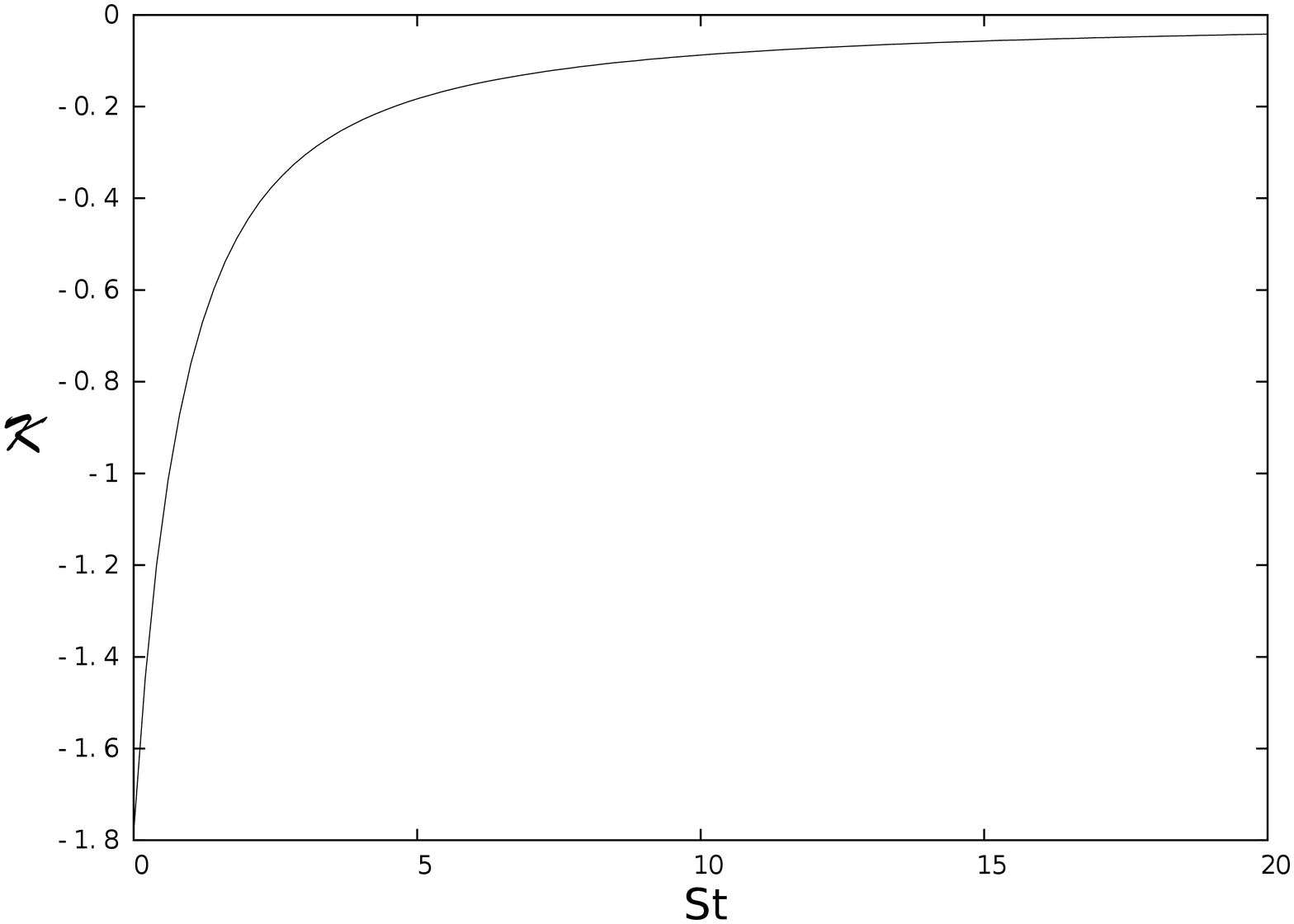}
\includegraphics[trim=0 0 0 0,clip,width=6cm, height=5cm]{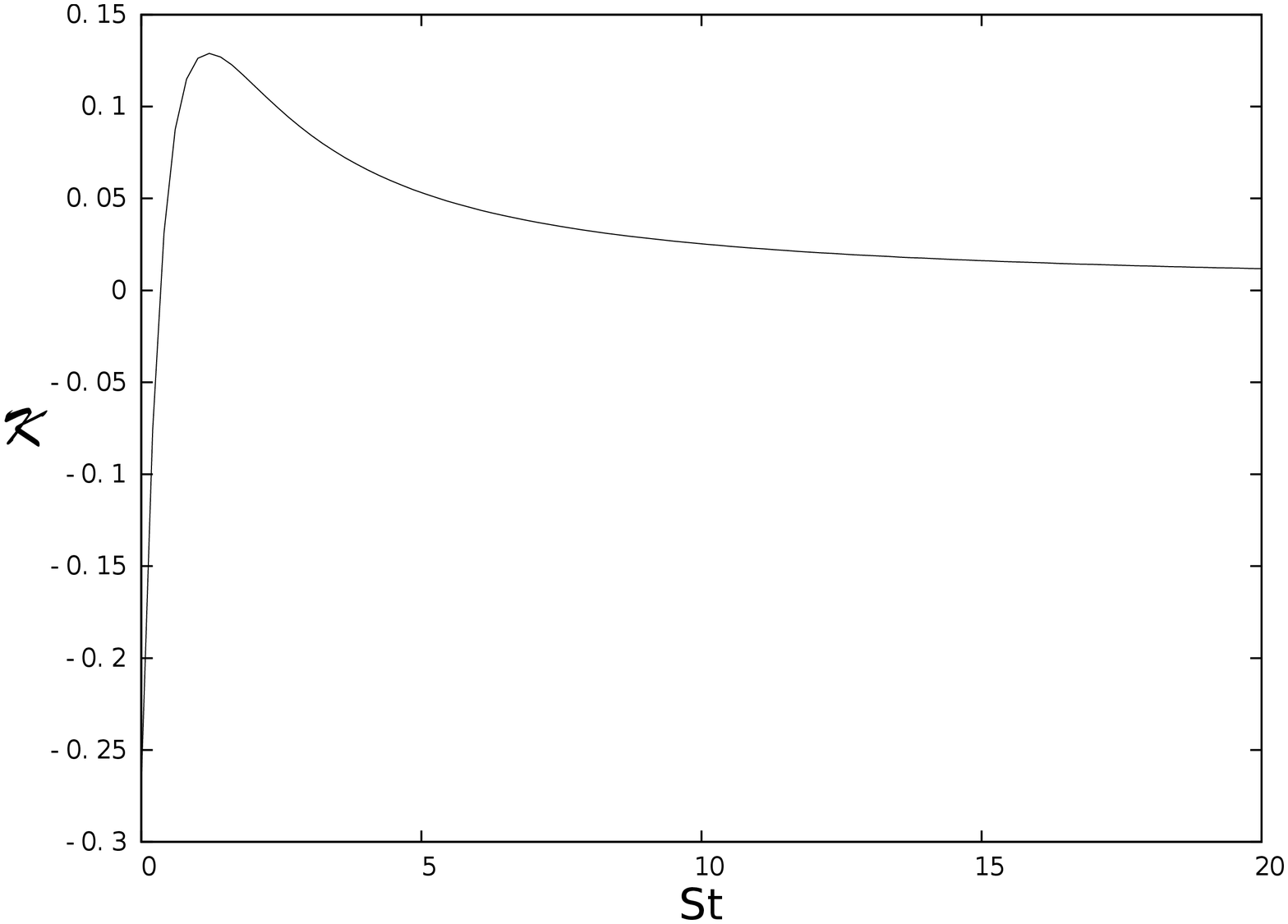}\\
\includegraphics[trim=0 0 0 0,clip,width=6cm, height=5cm]{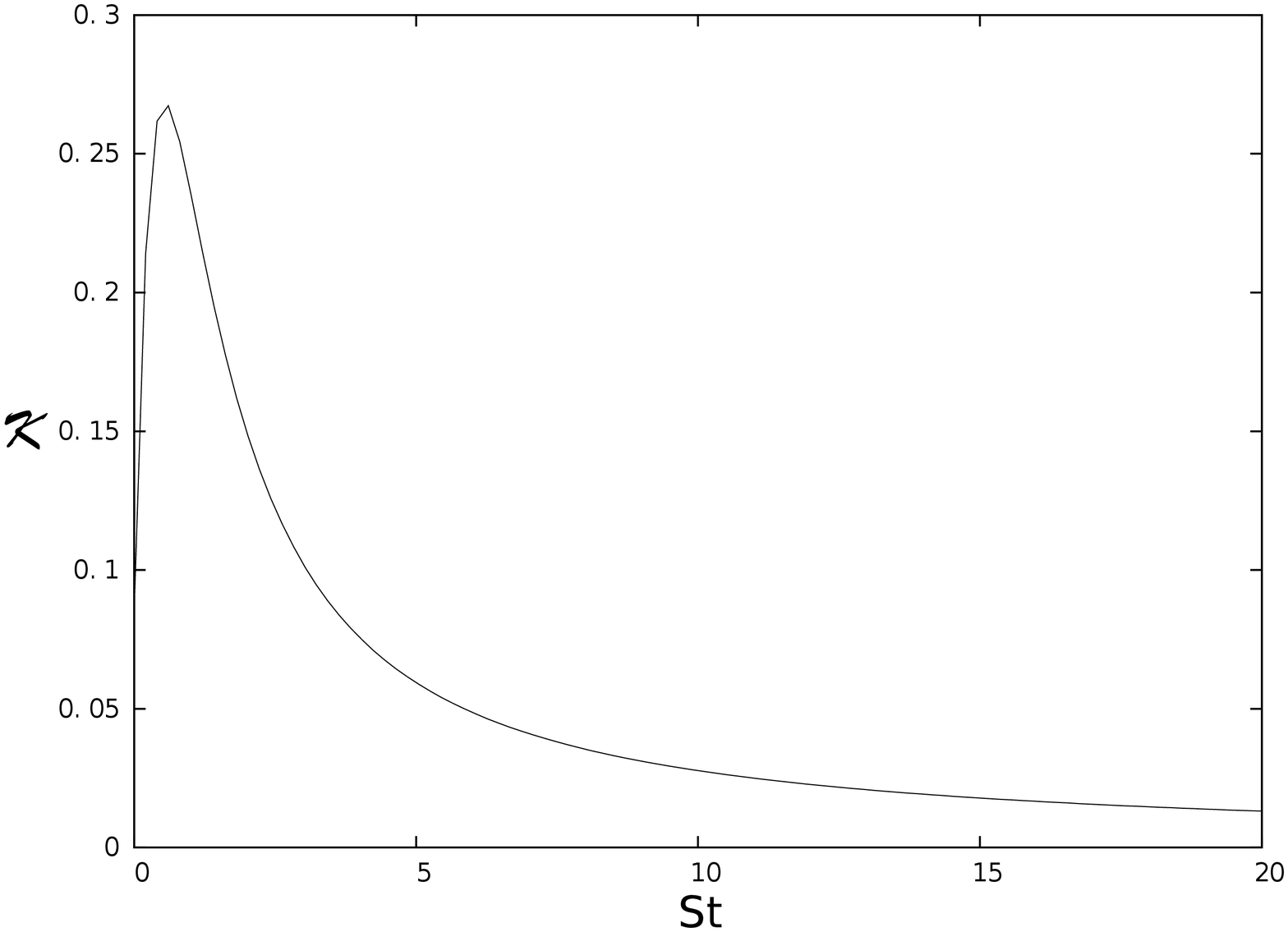}\\
\caption{$\mathcal K$ vs St at $\Omega=0.2$ (upper panel), $\Omega=0.8$ (middle panel) and $\Omega=1.1$ (lower panel).}
\label{fig1}
\end{figure}

The behavior of   $\mathcal K $ as a function of  $St$ is reported in Fig.~\ref{fig1} for different values of $\Omega$.  
For sufficiently small $\Omega$, $\mathcal K$  is negative and  inertia increases its value thus
enhancing transport.  For sufficiently large  $\Omega$, $\mathcal K$  is positive and inertia
increases its value up to a certain value of $St$ (corresponding to the intersection with the
dotted line of  Fig.~\ref{FigNuova})  above which transport is reduced by inertia.

The physical explanation of the resulting behavior of $\mathcal K$ vs  $St$, for small $St$,
can be traced back to the mechanism of transport enhancement
induced by a colored noise discussed in \cite{CaMa98}.
Indeed, the random contribution to the inertial particle velocity in (\ref{vel1}) turns out
to be a colored noise.
The fact that for large Stokes times $\mathcal K$ goes to zero is a simple consequence of the fact 
that in such a limit 
the contribution of the noise to the particle trajectories becomes negligible because of the large inertia of the particles.
 A maximum of transport is thus guaranteed in all cases where
$\mathcal K >0$ for $St=0$.

 In conclusion, by explicit computation, we have shown that the eddy diffusivities of
 inertial particles can be determined for the class of shear flows for all values of the Stokes
 number. Although the analysis has been here confined on the sole case of heavy particles,
 following the same line of reasoning it is not difficult to show that the present results actually
 hold for any density ratio of the particles (i.e. for any value of the added-mass term $\beta$
 involved in the model (2.2) of Ref.~\cite{mamamu2012}). 
 We also show that the analytical  results we obtained for the class of shear flows
 correspond to  the  leading order contribution either
 in the deviation from the shear flow geometry or in the P\'eclet  number of general
 random Gaussian velocity fields (i.e. not of shear type). \\  
 The results we obtained for the eddy-diffusivity allowed us to investigate the role of
 inertia on the asymptotic transport regime. It turned out that both enhancement and reduction of transport (with respect to the tracer case) may occur depending on the extension of
 anticorrelated regions  of the carrier flow Lagrangian auto-correlation function.

 AM acknowledges with thanks the financial support from the PRIN 2012 project n. D38C13000610001 funded by the Italian Ministry of Education. We are also grateful for the financial
support for the computational infrastructure from the Italian flagship project RITMARE.


\begin{thebibliography}{10}
\bibitem{IPCC}
IPCC: Fifth assessment report – the physical science basis, available at: http://www.ipcc.ch
(last access: 14 December 2015), (2013)

\bibitem{Denjean2016}
C. Denjean, F. Cassola, A. Mazzino, S. Triquet, S. Chevaillier, N. 
Grand, T. Bourrianne, G. Momboisse, K. Sellegri, A. Schwarzenbock, E. 
Freney, M. Mallet and P. Formenti,
Size distribution and optical properties of mineral dust aerosols 
transported in the western Mediterranean.
{\it Atmos. Chem. Phys.} {\bf 16}, 1081-1104 (2016).
 
\bibitem{martins2008}
M. Martins Afonso,
The terminal velocity of sedimenting particles in a flowing fluid.
{\it J. Phys. A: Math. Theor.} {\bf 41}, 385501 (2008)

  

\bibitem{frischeddy}
U. Frisch,
Lecture on turbulence and lattice gas hydrodynamics. In Lecture Notes, 
NCAR- GTP Summer School June 1987 (ed. J. R. Herring \& J. C. 
McWilliams), 219-371. World Scientific. (1987)

\bibitem{F95}
U. Frisch, Turbulence. Cambridge Univ. Press. (1995)


\bibitem{mamamu2012}
   M. Martins Afonso, A. Mazzino and P. Muratore-Ginanneschi,
   Eddy diffusivities of inertial particles under gravity, {\it J. 
Fluid Mech.} {\bf 694},
   426-463 (2012)




 

\bibitem{MR83}
M.R. Maxey and J.J. Riley, Equation of motion for a small rigid sphere in a nonuniform
flow. {\it Phys. Fluids} {\bf 26}, 883–889 (1983)

\bibitem{G83}
R. Gatignol, The Fax\'en formulae for a rigid particle in an unsteady non-uniform Stokes flow.
{\it J. M\'ec. Th\'eor. Appl.} {\bf 1}, 143–160 (1983)

\bibitem{Jacobs}
K. Jacobs, Stochastic Processes for Physicists. Understanding Noisy Systems. 
Cambridge Cambridge Univ. Press. (2010)

\bibitem{R88}
M.W. Reeks, The relationship between Brownian motion and the random motion of small particles in a
turbulent flow. {\it Phys. Fluids} {\bf 31}, 1314–1316 (1988)

\bibitem{MaCa99}
A. Mazzino and P. Castiglione, 
\newblock {Interference phenomena in scalar transport induced by a noise finite correlation time},
\newblock {\em Europhys. Lett. 45}, 476-481 (1999)

\bibitem{MazzVerg}
A. Mazzino and M. Vergassola, 
\newblock {Interference between turbulent and molecular diffusion},
\newblock {\em Europhys. Lett. 37}, 535-540 (1997)

  
\bibitem{Antonov}
N. V. Antonov and N. M. Gulitskiy, 
\newblock {Passive advection of
a vector field: Anisotropy, finite correlation time, exact
solution, and logarithmic corrections to ordinary scaling},
\newblock {\em Phys. Rev. E 92}, 043018 (2015)
 
\bibitem{Kaneda}
Y. Kaneda, T. Ishihara and K. Gotoh, 
\newblock {Taylor expansions
in powers of time of Lagrangian and Eulerian two-point
two-time velocity correlations in turbulence},
\newblock {\em Phys. Fluids 11}, 2154-2166 (1999)
 
 \bibitem{BoiMazzLac}
S. Boi,  A. Mazzino and G. Lacorata, 
\newblock {Explicit expressions
for eddy-diffusivity fields and effective large-scale advec-
tion in turbulent transport},
\newblock {\em J. Fluid Mech. 795}, 524-548 (2016)


\bibitem{dissip}
G. K. Batchelor, and A. A. Townsend, 
\newblock {Decay of Turbulence
in the Final Period},
\newblock {\em of the Royal Society A
194,} 527-543 (1948)

 \bibitem{CaMa98}
P. Castiglione and A. Mazzino, 
\newblock {Noise small correlation time effects on the dispersion of passive scalars},
\newblock {\em Europhys. Lett. 43}, 522-526  (1998)



\end{thebibliography}
\end{document}